\documentclass{svjour3}                     
\def\be{\begin{equation}}
\def\ee{\end{equation}}
\def\bea{\begin{eqnarray}}
\def\eea{\end{eqnarray}}
\def\btable{\begin{table}}
\def\etable{\end{table}}
\def\btabular{\begin{tabular}}
\def\etabular{\end{tabular}}
\def\bdoc{\begin{document}}
\def\edoc{\end{document}}
\def\bfig{\begin{figure}}
\def\efig{\end{figure}}
\def\ben{\begin{enumerate}}
\def\een{\end{enumerate}}
\def\bit{\begin{itemize}}
\def\eit{\end{itemize}}
\def\bitem{\begin{itemize}}
\def\eitem{\end{itemize}}
\def\bquote{\begin{quote}}
\def\equote{\end{quote}}

\def\bcenter{\begin{center}}
\def\ecenter{\end{center}}

\def\bsmall{\begin{small}}
\def\esmall{\end{small}}
\def\blarge{\begin{large}}
\def\elarge{\end{large}}
\def\bLarge{\begin{Large}}
\def\eLarge{\end{Large}}
\def\bhuge{\begin{huge}}
\def\ehuge{\end{huge}}
\def\bHuge{\begin{Huge}}
\def\eHuge{\end{Huge}}
\def\bfoot{\begin{footnotesize}}
\def\efoot{\end{footnotesize}}

\def\n{\noindent}
\def\bb{\left(}
\def\eb{\right)}
\def\bbs{\left[}
\def\ebs{\right]}
\def\bba{\left<}
\def\eba{\right>}
\def\bbsq{\left[}
\def\ebsq{\right]}
\def\bbcu{\left\{}
\def\ebcu{\right\}}
\def\bml{\left|}
\def\emr{\right|}

\def\etal{et al.\ }
\def\ie{i.\,e.\,, }
\def\eg{e.\,g.\,} 
\def\etc{etc.\ }
\def\viz{viz.\,, }

\def\gne #1#2{\ \vphantom{S}^{\raise-0.5pt\hbox{$\scriptstyle#1$}}_
{\raise0.5pt \hbox{$\scriptstyle#2$}}}
\def\gneq{\gne > \sim}
\def\lneq{\gne < \sim}

\def\mod #1{|#1|}
\def\vec#1{{\bf #1}}
\def\dotvec#1{\dot\vec{#1}}
\def\mvec#1{\mod{\vec{#1}}}
\def\mean #1{\bba #1 \eba}
\def\meanvsq#1{\mean{\vec{#1}^2}}
\def\dotp#1#2{\vec{#1}\cdot\vec{#2}}
\def\meandotp#1#2{\mean{\dotp{#1}{#2}}}
\def\crossp#1#2{\vec{#1}\times\vec{#2}}
\def\vecp{\vec{p}}
\def\vecpone{\vec{p}_{1}}
\def\vecn{\vec{n}}
\def\vecnone{\vec{n}_{1}}

\def\reci #1{\frac {1} {#1}}
\def\ten#1#2{#1{\times10^{#2}}}
\def\onlyten#1{10^{#1}}
\def\deg{\unit{deg}}
\def\degsq{\unit{deg^2}}
\def\ster{\unit{steradian}}
\def\perc{\unit{percent}}
\def\percent{\unit{percent}}

\def\deri#1#2{\frac {d#1} {d#2}}
\def\deriemp#1{\frac {d} {d#1}}
\def\pderi#1#2{\frac {\partial#1} {\partial#2}}
\def\pderiemp#1{\frac {\partial} {\partial#1}}
\def\secderi#1#2{\frac {d^2#1} {d#2^2}}
\def\secpderi#1#2{\frac {\partial^2#1} {\partial#2^2}}

\def\grad{\nabla}

\def\unit #1{\,{\rm #1}} 

\def\angst{\unit{\AA}}

\def\g{\unit{g}}
\def\gm{\unit{gm}}
\def\gmi{\gm^{-1}}
\def\mg{\unit{mg}}
\def\kg{\unit{kg}}
\def\nm{\unit{nm}}
\def\mum{\unit{\mu m}}
\def\mumi{\unit{\mu\,m^{-1}}}
\def\mm{\unit{mm}}
\def\cm{\unit{cm}}
\def\mtr{\unit{m}}
\def\mtri{\unit{m^{-1}}}
\def\km{\unit{km}}

\def\pc{\unit{pc}}
\def\kpc{\unit{kpc}}
\def\mpc{\unit{Mpc}}
\def\gpc{\unit{Gpc}}

\def\cmi{\unit{cm^{-1}}}
\def\cmsq{\unit{cm^2}}
\def\cmsqi{\unit{cm^{-2}}}
\def\cmcu{\unit{cm^3}}
\def\cmcui{\unit{cm^{-3}}}
\def\kmi{\unit{km^{-1}}}
\def\pci{\unit{pc^{-1}}}
\def\kpci{\unit{kpc^{-1}}}
\def\mtrsq{\unit{m^2}}
\def\mtrsqi{\unit{m^{-2}}}
\def\mtrcu{\unit{m^3}}
\def\mtrcui{\unit{m^{-3}}}
\def\mpccu{\unit{cm^3}}
\def\mpccui{\unit{cm^{-3}}}
\def\gpccu{\unit{cm^3}}
\def\gpccui{\unit{Gpc^{-3}}}
\def\kpsi{\unit{km \, s^{-1}}}

\def\ksec{\unit{ksec}}
\def\msec{\unit{msec}}
\def\musec{\unit{\mu\,sec}}
\def\min{\unit{min}}
\def\hr{\unit{hr}}
\def\day{\unit{day}}
\def\days{\unit{days}}
\def\week{\unit{week}}
\def\yr{\unit{yr}}
\def\myr{\unit{Myr}}
\def\gyr{\unit{Gyr}}
\def\micron{\unit{\mu m}}
\def\delto{\Delta t_{\rm o}}

\def\seci{\unit{s^{-1}}}
\def\dayi{\unit{day^{-1}}}
\def\yri{\unit{yr^{-1}}}

\def\cms{\unit{cm\,s^{-1}}}
\def\mts{\unit{m\,s^{-1}}}
\def\mtss{\unit{m\,s^{-2}}}
\def\kms{\unit{km\,s^{-1}}}
\def\kmh{\unit{km\,h^{-1}}}

\def\mas{\unit{mas}}
\def\muas{\unit{{\mu}}as}
\def\masyr{\unit{mas\,yr^{-1}}}
\def\arcs{\unit{arcsec}}
\def\arcsec{\unit{arcsec}}
\def\arcsecsq{\unit{arcsec^2}}
\def\arcsecsqi{\unit{arcsec^{-2}}}
\def\arcmin{\unit{arcmin}}
\def\arcm{\unit{arcmin}}
\def\arcmsq{\unit{arcmin^2}}
\def\degsq{\unit{deg^2}}
\def\degsqi{\unit{deg^{-2}}}

\def\vlos{\vec{v}_{\rm los}}

\def\hz{\unit{Hz}}
\def\khz{\unit{kHz}}
\def\mhz{\unit{MHz}}
\def\ghz{\unit{GHz}}
\def\thz{\unit{Thz}}
\def\ev{\unit{eV}}
\def\kev{\unit{keV}}
\def\mev{\unit{MeV }}
\def\gev{\unit{GeV }}
\def\tev{\unit{TeV }}

\def\kel{\unit{K}}

\def\newt{\unit{N}}

\def\funit{\rm\,erg\,cm^{-2}\,sec^{-1}}
\def\pfunit{\unit{cm^{-2}\,sec^{-1}}}
\def\funith{\rm\,erg\,cm^{-2}\,sec^{-1}\,Hz^{-1}}
\def\funitb{\rm\,erg\,cm^{-2}\,sec^{-1}\ster^{-1}}
\def\funitbh{\rm\,erg\,cm^{-2}\,sec^{-1}\Hz^{-1}\ster^{-1}}
\def\funitba{\rm\,erg\,cm^{-2}\,sec^{-1}\AA^{-1}\ster^{-1}}
\def\punitb{\rm\,photons\,cm^{-2}\,sec^{-1}\ster^{-1}}
\def\lunit{\rm\,erg\,sec^{-1}}
\def\lunith{\rm\,erg\,sec^{-1}\,Hz^{-1}}
\def\whz{\unit{W\,Hz^{-1}}}
\def\whzs{\unit{W\,Hz^{-1}\,str^{-1}}}
\def\wmtrsqi{\unit{W\,m^{-2}}}
\def\jy{\unit{Jy}}
\def\mjy{\unit{mJy}}
\def\mujy{\unit{\mu Jy}}
\def\jyu{\unit{W\,m^{-2}\,Hz}}
\def\erg{\unit{erg}}
\def\ergs{\unit{erg\,s^{-1}}}
\def\ergsh{\unit{erg\,s^{-1}\,Hz^{-1}}}
\def\mcrab{\unit{milliCrab}}
\def\ergsc{\unit{erg\,s^{-1}\,cm^{-3}}}
\def\ergcmcu{\unit{erg\,cm^{-3}}}
\def\ergscmcu{\unit{erg\,s^{-1}\,cm^{-3}}} 
\def\ergcmt{\unit{erg\,cm^{-3}}}
\def\ergcmcui{\unit{erg\,cm^{-3}}}

\def\gmcmcui{\gm\cmcui}
\def\kgmtrcui{\kg\mtrcui}
\def\gcmcui{\g\cmcui}
  
\def\melec{m_{\rm e}}
\def\mectwo{\melec c^2}
\def\mprot{m_{\rm p}}
\def\nelec{n_{\rm e}}
\def\nion{n_{\rm i}}
\def\nprot{n_{\rm p}}
\def\pelec{p_{\rm e}}
\def\pfermi{p_{\rm f}}
\def\efermi{E_{\rm f}}
\def\mue{\mu_{\rm e}}

\def\gntff{\overline{g}_{\rm ff}}

\def\eneutrino{\nu_{\rm e}}
\def\eaneutrino{\overline{\nu}_{\rm e}}
\def\mneutrino{\nu_{\mu}}
\def\maneutrino{\\overline{nu}_{\mu}}
\def\tneutrino{\nu_{\tau}}
\def\maneutrino{\\overline{nu}_{\tau}}

\def\fine{\alpha_{\rm f}}
\def\comptw{\lambda_{\rm e}}

\def\threehe{^3{\rm He}}
\def\fourhe{^4{\rm He}}
\def\sixli{^6{\rm Li}}
\def\sevenli{^7{\rm Li}}
\def\ninebe{^9{\rm Be}}
\def\tenb{^{10}{\rm B}}
\def\elevenb{^{11}{\rm B}}
\def\twelvec{^{12}{\rm C}}
\def\sixteeno{^{16}{\rm O}}

\def\fourhebyh{\fourhe/{\rm H}}
\def\threehebyh{\threehe/{\rm H}}
\def\sevenlibyh{^7{\rm Li}/{\rm H}}
\def\bsqfh{{\rm B^2FH}}

\def\mag{\unit{magnitude}}
\def\mags{\unit{magnitudes}}
\def\absm#1{M_{\rm#1}}
\def\appm#1{m_{\rm#1}}
\def\Delm#1{\Delta m_{#1}}
\def\magarcsecsqi{\unit{mag\arcsecsqi}}

\def\lsol{L_{\odot}}
\def\msol{M_{\odot}}
\def\rsol{R_{\odot}}
\def\zsol{Z_{\odot}}

\def\aunit{\unit{AU}}

\def\mear{M_{\oplus}}
\def\rear{R_{\oplus}}
\def\aear{a_{\oplus}}
\def\pear{P_{\oplus}}

\def\mjup{M_{\rm Jupiter}}
\def\ajup{a_{\rm Jupiter}}
\def\pjup{P_{\rm Jupiter}}
\def\msat{M_{\rm Saturn}}

\def\amer{a_{\rm Mercury}}
\def\pmer{P_{\rm Mercury}}

\def\aplu{a_{\rm Pluto}}
\def\pplu{P_{\rm Pluto}}

\def\meight{\bb\frac{M}{\onlyten{8}\msol}\eb}

\def\mbymsol{\bb\frac{M}{\msol}\eb}

\def\aop{\alpha_{\rm op}}
\def\ar{\alpha_{\rm R}}
\def\ax{\alpha_{\rm x}}
\def\ag{\alpha_{\gamma}}
\def\aire{\alpha_{2.2-25\micron}}
\def\asm{\alpha_{\rm sm}}
\def\onemal{1-\alpha}
\def\nuone{\nu_1}
\def\nutwo{\nu_2}
\def\nuz{\nu_0}
\def\nug{\nu_g}
\def\nuc{\nu_c}
\def\nua{\nu_a}
\def\nup{\nu_p}
\def\nux{\nu_{\rm x}}
\def\nuop{\nu_{\rm op}}
\def\nupz{\nu_{p0}}
\def\nuaop{\nu^{-\alpha_{op}}}
\def\nualpha{\nu^{-\alpha}}
\def\alnu{\alpha_{\nu}}
\def\alnua{\alpha_{\nu_a}}
\def\alnusc{\alnu^{\rm sc}}
\def\jnu{j_{\nu}}
\def\jnua{j_{\nu_a}}
\def\tstar{\tau_{*}}
\def\tnu{\tau_{\nu}}
\def\etnu{e^{\tnu}}
\def\emtnu{e^{-\tnu}}
\def\fnu{F(\nu)}
\def\fnua{F(\nua)}
\def\fr{F_{\rm R}}
\def\fop{F_{\rm op}}
\def\fopl{F_{\rm op}^l}
\def\fx{F_{\rm x}}
\def\fxl{F_{\rm x}^{l}}
\def\frlim{F_{R,lim}}
\def\mv{m_V}
\def\mb{m_B}
\def\mr{m_R}
\def\Mv{M_V}
\def\Mb{M_B}
\def\Mr{M_R}

\def\bnu{B(\nu)}
\def\blnu{B_{\nu}}
\def\inu{I(\nu)}
\def\ilnu{I_{\nu}}
\def\snu{S(\nu)}
\def\slnu{S_{\nu}}
\def\Snu{S(\nu)}
\def\unu{u(\nu)}
\def\ulnu{u_{\nu}}
\def\Snua{S(\nu_a)}
\def\nnu{n(\nu)}
\def\nlnu{n_{\nu}}
\def\nomega{n(\omega)}
\def\nlomega{n_{\omega}}
\def\taunu{\tau_{\nu}}
\def\etaunu{e^{\taunu}}
\def\emtaunu{e^{-\taunu}}
\def\exptaunu{\exp(\taunu)}
\def\expmtaunu{\exp(-\taunu)}
\def\planckinu{\frac{2h\nu^3/c^2}{\exp(h\nu/kT)-1}}
\def\planckunu{\frac{8\pi h\nu^3/c^3}{\exp(h\nu/kT)-1}}
\def\planckilambda{\frac{2hc^2/\lambda^5}{\exp(hc/\lambda kT)-1}}
\def\planckulambda{\frac{8\pi hc/\lambda^5}{\exp(hc/\lambda kT)-1}} 

\def\plisnu{\frac{2h\nu^3/c^2}{\exp(h\nu/kT)-1}}
\def\plusnu{\frac{8\pi h\nu^3/c^3}{\exp(h\nu/kT)-1}}
\def\plislambda{\frac{2hc^2/\lambda^5}{\exp(hc/\lambda kT)-1}}
\def\pluslambda{\frac{8\pi hc/\lambda^5}{\exp(hc/\lambda kT)-1}} 
\def\plnut{\frac{2h\nu^3/c^2}{\exp(h\nu/kT)-1}}
\def\rjnut{\frac{2kT\nu^2}{c^2}}

\def\Pnu{P(\nu)}
\def\lnu{L(\nu)}
\def\llnu{\log \lnu}
\def\lr{L_{\rm R}}
\def\llr{\log L_{\rm R}}
\def\lrlim{L_{R,lim}}
\def\lfive{L(5\ghz)}
\def\lop{L_{{\rm op}}}
\def\llop{\log L_{{\rm op}}}
\def\lx{L_{\rm x}}
\def\llx{\log L_{\rm x}}
\def\lxmin{L_{\rm x,min}}
\def\lxmax{L_{\rm x,max}}
\def\lmax{L_{|rm max}}
\def\lir{L_{\rm IR}}
\def\luv{L_{\rm UV}} 
\def\lrc{L_{\rm rc}}
\def\llrc{\log L_{\rm rc}}
\def\lrext{L_{\rm \rm r,ext}}
\def\llrext{\log L_{r,ext}}
\def\lre{L_{\rm re}}
\def\llre{\log L_{\rm re}}
\def\lxb{L_{\rm xb}}
\def\llxb{\log L_{\rm xb}}
\def\lxu{L_{\rm xu}}
\def\llxu{\log L_{\rm xu}}
\def\lre{L_{\rm re}}
\def\llre{\log L_{\rm re}}
\def\rt{R_{\rm t}}
\def\rx{R_{\rm x}}
\def\rxbu{R_{\rm xbu}}
\def\rtx{R_{\rm tx}}
\def\gxbp{g_{\rm x}(\beta, \psi)}
\def\grbp{g_{\rm r}(\beta, \psi)}
\def\aox{\alpha_{\rm ox}}
\def\aoxl{\alpha_{\rm ox}^l}
\def\aro{\alpha_{\rm ro}}
\def\arx{\alpha_{\rm rx}}
\def\aoxx{\alpha_{\rm oxx}}
\def\auv{\alpha_{\rm uv}}
\def\sigt{\sigma_{\rm T}}
\def\sigkn{\sigma_{\rm KN}}
\def\signu{\sigma_{\nu}}
\def\siggg{\sigma_{\gamma\gamma}}
\def\tausc{\tau_{\rm sc}}
\def\taut{\tau_{\rm T}}
\def\nsc{N_{\rm sc}}
\def\freenu{l_{\nu}}
\def\U #1{U_{\rm #1}}
\def\eps{\epsilon}
\def\epsone{\epsilon_1}
\def\epstwo{\epsilon_2}
\def\epsx{\epsilon_{\rm x}}
\def\epsxs{\eps_{\rm xs}}
\def\epss{\epsilon_{\rm s}}
\def\epsg{\epsilon_{\gamma}}
\def\epsnu{\epsilon_{\nu}}
\def\epsir{\eps_{\rm IR}}
\def\epsi{\eps_{\rm i}}
\def\epsf{\eps_{\rm f}}
\def\DOM{\Delta\Omega}
\def\omegaone{\omega_{1}}

\def\hnu{h\nu}
\def\hnubyc{\frac{h\nu}{c}}
\def\hbarom{\hbar\omega}

\def\lcapc{L_{\rm C}}
\def\lcaps{L_{\rm S}}
\def\lcapi{L{\rm i}}
\def\lcaph{L_{\rm h}}
\def\dtvar{\Delta t_{\rm var}}
\def\tbri{T_{\rm b}}
\def\tsri{T_{\rm i}}
\def\Tvar{T_{\rm var}}
\def\dmv{dm_{\rm V}}
\def\betaa{\beta_{\rm a}}
\def\vela{v_{\rm a}}
\def\te{t_{\rm e}}
\def\dte{dt_{\rm e}}
\def\gammap{\gamma_{\rm p}}
\def\gammab{\gamma_{\rm b}}
\def\gammai{\gamma{\rm i}}
\def\gammaj{\gamma{\rm j}}
\def\gammac{\gamma{\rm c}}

\def\vbyvm{\frac {V} {V_m}}
\def\vbyvmp{\frac {V'} {V_m'}}
\def\mvbyvm{\bba\vbyvm\eba}
\def\mvbyvmp{\bba\vbyvmp\eba}

\def\vbyvml{V/V_m}
\def\vbyvmpl{V'/V_m'}
\def\mvbyvml{\bba\vbyvml\eba}
\def\mvbyvmpl{\bba\vbyvmpl\eba}

\def\bperp{B_{\perp}}
\def\bpar{B_{\parallel}}
\def\bperppar{B_{\perp-\parallel}}
\def\dtvar{\Delta t_{var}}
\def\Tvar{T_{var}}
\def\dmv{dm_v}
\def\betaa{\beta_a}
\def\gammap{\gamma_{\rm p}}
\def\gammab{\gamma_{\rm b}}
\def\gammamax{\gamma_{\rm max}}
\def\scrid{{\cal D}}
\def\scril{{\cal L}}
\def\scrio{{\cal O}}
\def\scrir{{\cal R}}
\def\fbeam{F_b}
\def\fext{F_{ext}}
\def\fin{F_{\rm in}}
\def\fout{F_{\rm out}}
\def\bcospsi{\beta\cos\psi}
\def\bsinpsi{\beta\sin\psi}
\def\rnu{R_{\nu}}
\def\rnuone{R_{\nu_1}}
\def\rnutwo{R_{\nu_2}}

\def\nh{N_{\rm H}}
\def\mh{M_{\rm H}}
\def\eax{E^{-\ax}}
\def\ex{E_{\rm x}}
\def\ex{E_{\rm xs}}

\def\zmin{z_{\rm min}}
\def\zmax{z_{\rm max}}
\def\rmin{r_{\rm min}}
\def\rmax{r_{\rm max}}
\def\rsc{r_{\rm sc}}
\def\rscm{r_{\rm sc,m}}

\def\corrxz{r_{\rm x,z}}
\def\corrxop{r_{\rm x,op}}
\def\corropz{r_{\rm op,z}}
\def\corrxopz{r_{\rm x,op;z}}
\def\chisq{\chi^2}
\def\chisqr{\chi^{2}_{\nu}}

\def\philx{\Phi{(\lx)}}
\def\phixstar{\Phi_{X}^{*}}
\def\phixstaro{\Phi_{X1}^{*}}
\def\lxstar{\lx^{*}}
\def\lxfour{L_{{\rm x},44}}
\def\lxstarfour{L_{{\rm x},44}^*(0)}

\def\asca{{\em ASCA }}
\def\axaf{{\em AXAF }}
\def\ein{{\em EINSTEIN }}
\def\exosat{{\em EXOSAT }}
\def\ginga{{\em GINGA }}
\def\rosat{{\em ROSAT }}
\def\compton{{\em COMPTON }}
\def\granat{{\em GRANAT }}
\def\iras{{\em IRAS }}
\def\iue{{\em IUE }}

\def\seyone{Seyfert\,1 }
\def\seytwo{Seyfert\,2 }
\def\bllac{BL Lac }
\def\bllacs{BL Lacs }
\def\rbl{{\rm RBL }}
\def\xbl{{\rm XBL }}
\def\ipc{{\rm IPC }}
\def\pspc{{\rm PSPC }}

\def\cfour{{\rm C\,IV }}
\def\cthreesf{{\rm C\,III] }}
\def\fetwo{{\rm Fe\,II }}
\def\halpha{{\rm H\,\alpha }}
\def\hbeta{{\rm H\,\beta }}
\def\hgamma{{\rm H\,\gamma}}
\def\hi{{\rm H\,I }}
\def\lyal{{\rm Ly\,\alpha }}
\def\mgtwo{{\rm Mg\,II }}
\def\ntwo{{\rm [N\,II]}}
\def\nfive{{\rm N\,V }}
\def\otwo{{\rm [O\,II] }}
\def\othree{{\rm [O\,III] }}
\def\kalpha{K_{\alpha} }
\def\ykz{Y^{K}_{Z} }

\def\const{{\rm constant}}
\def\gray{\gamma-{\rm ray}}
\def\grays{\gamma-{\rm rays}}
\def\maxtau{\tausc(1+\tausc)}
\def\maxtnu{\tnu(1+\tnu)}
\def\dm{{\rm DM}}
\def\dmsinmb{\dm\sin\mod{b}}
\def\dmsinb{\dm\sin b}
\def\sbyn{\frac{S}{N}}

\def\noi{\noindent}
\def\vf{\vspace{0.5cm}}
\def\vff{\vspace{1cm}}
\def\pbr{\pagebreak}
\def\pnew{\newpage}
\def\pemp{\pagestyle{empty}}
\def\tpemp{\thispagestyle{empty}}

\newcommand{\onlytitle}[1]
{\title{#1}
\author{ }
\date{ }
\maketitle\tpemp }

\newcommand{\onlyfig}[2]
{\begin{center}\epsfig{figure=#1.ps, width=#2in}\end{center}}

\newcommand{\onlyfige}[2]
{\begin{center}\epsfig{figure=#1.eps, width=#2in}\end{center}}

\def\vbyc{\frac{v}{c}}
\def\vbycsq{\frac{v^2}{c^2}}
\def\omvbycsq{\sqrt{1-\vbycsq}}

\def\gmr{\frac{GM}{r}}
\def\gmcsr{\frac{GM}{c^2r}}
\def\gmcsrl{GM/c^2r}
\def\tgmcsr{\frac{2GM}{c^2r}}
\def\schwarz{r_{\rm S}}
\def\tschwarz{t_{\rm S}}
\def\rergo{r_{\rm E}}
\def\rplus{r_{+}}
\def\mbh{M_{\rm BH}}
\def\mbyl{M/L}

\def\lh{l_{\rm h}}
\def\li{l_{\rm i}}
\def\ls{l_{\rm s}}

\def\omk{\Omega_{\rm K}(r)}
\def\velr{v_{r}}
\def\velp{v_{\phi}}
\def\velz{v_{z}}
\def\sigrt{\Sigma(r,t)}
\def\mach{{\cal M}}
\def\scrir{{\cal R}}
\def\tin{t_{\rm in}}
\def\tl{t_{\rm L}}
\def\ldisk{L_{\rm disk}}
\def\temps{T_{\rm s}}
\def\tempc{T_{\rm c}}
\def\tempb{T_{\rm B}}
\def\bnu{B_{\nu}}
\def\bnuts{B_{\nu}(\temps)}
\def\bnutsr{B_{\nu}(\temps(r))}

\def\rout{r_{\rm out}}
\def\tvis{t_{\rm vis}}
\def\pgas{p_{\rm g}}
\def\prad{p_{\rm r}}
\def\csou{c_{\rm s}}
\def\rstar{r_{*}}
\def\ledd{L_{\rm Edd}}
\def\mdedd{\dot{M}_{\rm Edd}}
\def\tedd{t_{\rm Edd}}
\def\tempedd{T_{\rm Edd}}
\def\bedd{B_{\rm Edd}}
\def\mdot{{\dot M}}
\def\geff{\vec{g}_{\rm eff}}
\def\phir{\Phi_{\rm rot}}
\def\phie{\Phi_{\rm eff}}
\def\zsr{z_{\rm s}(r)}
\def\udzero{u_0}
\def\udphi{u_{\phi}}
\def\uuzero{u^0}
\def\uuphi{u^{\phi}}

\def\isnu{I_{\nu}}
\def\islambda{I_{\lambda}}
\def\fsnu{F_{\nu}}
\def\fslambda{F_{\lambda}}

\def\plisnu{\frac{2h\nu^3/c^2}{\exp(h\nu/kT)-1}}
\def\plusnu{\frac{8\pi h\nu^3/c^3}{\exp(h\nu/kT)-1}}
\def\plislambda{\frac{2hc^2/\lambda^5}{\exp(hc/\lambda kT)-1}}
\def\pluslambda{\frac{8\pi hc/\lambda^5}{\exp(hc/\lambda kT)-1}} 

\def\isnurj{\isnu^{\rm RJ}}
\def\isnuw{\isnu^{\rm W}}

\def\cv{C_V}
\def\cp{C_p}
\def\dqdtv{\bb\pderi{Q}{T}\eb_V}
\def\dqdtp{\bb\pderi{Q}{T}\eb_p}

\def\CH{{\it CH}}

\def\hub{H_0}
\def\hubi{H^{-1}_0}
\def\hubble#1{H_0={#1}\unit{km\,sec^{-1}\,Mpc^{-1}}}
\def\hubbleunit{\unit{km\,sec^{-1}\,Mpc^{-1}}}
\def\hubhund{h_{100}}
\def\hubhundcu{\hubhund^3}
\def\hubhundcui{\hubhund^{-3}}
\def\hubhundi{\hubhund^{-1}}
\def\hubhundis{\hubhund^{-2}}
\def\cuponh{\bb \frac {c} {H_0} \eb}

\def\qz{q_0}
\def\qzsq{\qz^2}
\def\fqz{\bbcu\reci{\qzsq}\bbsq \qz z + (q_0-1)(\sqrt{1+2q_0z}-1)
\ebsq\ebcu}
\def\fqzno{\qz z + (q_0-1)(\sqrt{1+2q_0z}-1)}
\def\omb{\Omega_{{\rm b}}}
\def\omhi{\Omega_{{\rm H\,I}}}
\def\omlls{\Omega_{{\rm LLS}}}
\def\omlb{\Omega_{{\rm lb}}}
\def\omigm{\Omega_{{\rm IGM}}}
\def\omdlyal{\Omega_{{\rm DLy\,\alpha}}}
\def\zpr{z^{\prime}}
\def\zabs{z_{{\rm abs}}}
\def\zem{z_{{\rm em}}}
\def\zlim{z_{\rm lim}}
\def\zlimg{z_{\rm lim,G}}
\def\zlimgq{z_{\rm lim,GQ}}
\def\dlum{D_{\rm L}}
\def\dlumsq{D^2_{\rm L}}
\def\dmet{d_{\rm m}}
\def\dang{d_{\rm A}}

\def\lamo#1{\lambda_{\rm o_{#1}}}
\def\lame#1{\lambda_{\rm e_{#1}}}

\def\re{r_{\rm e}}
\def\rbyre{\frac{r}{\re}}
\def\rbyredev{\bb\rbyre\eb^{1/4}}
\def\rbyren{\bb\rbyre\eb^{1/n}}
\def\rbyrel{r/\re}
\def\rbyredevl{(\rbyrel)^{1/4}}
\def\rbyrenl{(\rbyrel)^{1/n}}
\def\rebyrd{\frac{\re}{\rd}}
\def\rebyrdl{\re/\rd}

\def\izero{I(0)}
\def\ire{I(\re)}
\def\ir{I(r)}
\def\icr{I_{\rm c}(r)}

\def\muzero{\mu(0)}
\def\mure{\mu(\re)}
\def\mur{\mu(r)}

\def\ibzero{I_{\rm b}(0)}
\def\ibre{I_{\rm b}(\re)}
\def\mibre{\mean{\ibre}}
\def\iblre{I_{\rm b}(<\re)}
\def\ibr{I_{\rm b}(r)}
\def\iblr{I_{\rm b}(<r)}
\def\ibtot{I_{\rm b,tot}}

\def\mubzero{\mu_{\rm b}(0)}
\def\mubre{\mu_{\rm b}(<\re)}
\def\mmubre{\mean{\mubre}}
\def\mubr{\mu_{\rm b}(r)}
\def\mulim{\mu_{\rm lim}}

\def\mudzero{\mu_{\rm d}(0)}
\def\mugzero{\mu_{\rm G}(0)}

\def\rd{r_{\rm d}}
\def\rbyrd{\frac{r}{\rd}}
\def\rbyrdl{r/\rd}
\def\idzero{I_{\rm d}(0)}
\def\idr{I_{\rm d}(r)}
\def\idlr{I_{\rm d}(<r)}
\def\idtot{I_{\rm d,tot}}

\def\imax{I_{\rm max}}
\def\imin{I_{\rm min}}

\def\dev{\onlyten{-3.33\rbyredevl}}
\def\devn{\onlyten{-\cn\rbyren}}
\def\exprbyrd{e^{-(\rbyrdl)}}
\def\cn{c_n}

\def\dbyb{\frac{D}{B}}
\def\bbyd{\frac{B}{D}}
\def\dbybl{D/B}
\def\bbydl{B/D}

\def\absgal{M_{\rm G}}
\def\absbul{M_{\rm B}}
\def\absdisk{M_{\rm d}}
\def\absq{M_{\rm Q}}

\def\lstar{L_{*}}

\def\mbyl{\bb\frac{M}{L}\eb}

\def\labchap #1{\label{chap:#1}}
\def\labequn #1{\label{eq:#1}}
\def\labfig #1{\label{fig:#1}}
\def\labsecn #1{\label{sec:#1}}
\def\labsubsecn #1{\label{subsecn:#1}}
\def\labsubsubsecn #1{\label{subsubsecn:#1}}
\def\labtablem #1{\label{tab:#1}}
\def\labapp #1{\label{app:#1}}
\def\labboxl #1{\label{box:#1}}

\def\boxl #1{box~\ref{box:#1}}
\def\chap #1{Chapter~\ref{chap:#1}}
\def\equn #1{Equation~\ref{eq:#1}}
\def\fig #1{Figure~\ref{fig:#1}}
\def\relation #1{relation~\ref{eq:#1}}
\def\secn #1{Section~\ref{sec:#1}}
\def\subsecn #1{Section~\ref{subsecn:#1}}
\def\subsubsecn #1{Section~\ref{subsubsecn:#1}}
\def\app#1{Appendix~\ref{app:#1}}
\def\tablem #1{Table~\ref{tab:#1}}

\def\dequn#1#2{Equations~{\ref{eq:#1}}~and~{\ref{eq:#2}}}
\def\tequn#1#2#3{Equations~{\ref{eq:#1}},~{\ref{eq:#2}} and ~{\ref{eq:#3}}}
\def\mequn#1#2{Equations~{\ref{eq:#1}}~to~{\ref{eq:#2}}}
\def\dfig#1#2{Figures~{\ref{fig:#1}}~and~{\ref{fig:#2}}}
\def\mfig#1#2{Figures~{\ref{fig:#1}}~to~{\ref{fig:#2}}}
\def\dtablem#1#2{Tables~{\ref{tab:#1}}~and~{\ref{tab:#2}}}
\def\dsubsecn#1#2{Subsections~{\ref{subsecn:#1}}~and~{\ref{subsecn:#2}}}

\def\tobs#1{{\it The Observatory}{\bf #1}}

\def\cfa#1{{\it Center for Astrophysics Preprint }{\bf #1}}
\def\asp{Astron. Soc. of the Pacific, San Francisco.}
\def\clar{Clarendon Press, Oxford.} 
\def\cup{Cambridge University Press, Cambridge.}
\def\free{W. H. Freeman, New York.}
\def\klu{Kluwer, Dordrecht.}
\def\nrao{NRAO, Greenbank.}
\def\plenumn{PLenum Press, New York.}
\def\prince{Princeton University Press, Princeston}
\def\ridel{Ridel, Dordrecht.}
\def\springerb{Springer-Verlag, Berlin.}
\def\springern{Springer-Verlag, New York.}
\def\wiley{Wiley, New York.}
\def\yale{Yale University Press, New Haven.}

\def\noin{\noindent}

\def\sqr#1#2{{\vcenter{\vbox{\hrule height.#2pt
  \hbox {\vrule width.#2pt height#1pt \kern#1pt
  \vrule width.#2pt}
  \hrule height.#2pt}}}}

\def\square{\mathchoice\sqr68\sqr68\sqr{2.1}3\sqr{1.5}3}

\def\CH{{\it CH }}

\newcommand{\myruleb}[1]{\vspace{0.5cm}\noindent\rule{#1}{0.05in}}
\newcommand{\myrulee}[1]{\newline\noindent\rule{#1}{0.05in\vspace{0.5cm}}}

\newenvironment{mybox}[3]
{\begin{minipage}{4.5in}
\setlength{\parindent}{0.5cm}
\myruleb{4.5in}
\label{box:#1}
\begin{center}\fbox{\textbf{#2}}\end{center}
{\begin{small}#3\end{small}}
\myrulee{4.5in}
\end{minipage}}

\newenvironment{myboxp}[4]
{\begin{minipage}{4.5in}
\setlength{\parindent}{0.5cm}
\myruleb{4.5in}
\label{box:#1}
\begin{center}\fbox{\textbf{#2}}\end{center}
\begin{center}\epsfig{figure=#3.ps,width=2in}\end{center}
{\begin{small}#4\end{small}}
\myrulee{4.5in}
\end{minipage}}

\newcommand{\myfig}[2]
{\begin{figure}[ht]
\begin{center}
\label{fig:#1}
\epsfig{figure=#1.ps,width=5in}
\caption{#2}
\end{center}
\end{figure}}

\newcommand{\myplotone}[2]
{\begin{figure}[ht]
\begin{center}
\label{fig:#1}
\plotone{#1.ps}
\caption{#2}
\end{center}
\end{figure}}

\newcommand{\myequn}[2]
{\begin{equation}
{#2}
\labequn{#1}
\end{equation}}

\newcommand{\myequna}[2]
{\begin{eqnarray}
{#2}
\labequn{#1}
\end{eqnarray}}

\newenvironment{headnoff}[1]
{\pnew\bHuge\bcenter{\bf\underline{#1}}\ecenter\eHuge}

\newcommand{\state}[1]{\noindent{#1}\vf}

\newcommand{\mnote}[1]{{{\small\it\marginpar{#1}}}}

\def\cmcu{\unit{cm^3}}
\def\cmcui{\unit{cm^{-3}}}
\def\msol{M_{\odot}}
\def\cmcui{\unit{cm^{-3}}}
\def\gcmcui{\g\cmcui}
\def\gmcmcui{\gm\cmcui}
\smartqed  
\usepackage{graphicx}
\begin{document}

\title{Can a Black Hole Collapse to a Space-time Singularity?
}


\author{R. K. Thakur         
}


\institute{R. K. Thakur \at
  Retired Professor of Physics,\\
   School   of   Studies in Physics,\\
   Pt. Ravishankar Shukla University,
   Raipur, India\\
   Tel.: +91-771-2255168\\
   \email{rkthakur0516@yahoo.com}           
}

\date{Received: date / Accepted: date}

\maketitle

\begin{abstract}
A critique of the singularity theorems of Penrose, Hawking, and Geroch is given.
It is pointed out that a gravitationally collapsing black hole acts as an ultrahigh energy 
particle accelerator that can accelerate particles to energies inconceivable in any
terrestrial particle accelerator, and that when the energy $E$ of the particles
comprising matter in a black hole is $\sim 10^{2} GeV$ or more, or equivalently,
the temperature $T$ is $\sim 10^{15} K$ or more, the entire matter in the black hole
is converted into quark-gluon plasma permeated by leptons. As quarks and leptons
are fermions, it is emphasized that the collapse of a black-hole to a space-time
singularity is inhibited by Pauli's exclusion principle. It is also suggested that
ultimately a black hole may end up either as a stable quark star, or as a pulsating
quark star which may be a source of gravitational radiation, or it may simply explode
with a mini bang of a sort.
\end{abstract}
\section{Introduction}
When all the thermonuclear sources of energy of a star are exhausted, the core
of the star begins to contract gravitationally because, practically, there is no
radiation pressure to arrest the contraction, the pressure of matter being
inadequate for this purpose. If the mass of the core is less than the
Chandrasekhar limit ($\sim 1.44 \msol$), the contraction stops when the density of
matter in the core,  $\rho > 2 \times  10^{6} \gcmcui$; at this stage the
pressure of the
relativistically degenerate electron gas in the core is enough to withstand the
force of gravitation. When this happens, the core becomes a stable white dwarf.
However, when the mass of the core is greater than the Chandrasekhar limit, the
pressure of the relativistically degenerate electron gas is no longer sufficient
to arrest the gravitational contraction, the core continues to contract and becomes
denser and denser; and when the density reaches the value $\rho \sim 10^{7}\gcmcui$,
the process of neutronization sets in; electrons and protons in the core begin to
combine into neutrons through the reaction

\begin{center}
$p + e^{-} \rightarrow n + \nu_{e}$
\end{center}

The electron  neutrinos $\nu_{e}$ so produced escape from the core of the star. The gravitational
contraction continues and eventually, when the density of the core reaches the
value $\rho \sim 10^{14} \gcmcui$, the core consists almost entirely of
neutrons. If the mass of the core is less than the Oppenheimer-Volkoff limit ($\sim 3\msol$),
then at this stage the contraction stops; the pressure of the
degenerate neutron gas is enough to withstand the gravitational force. When this
happens, the core becomes a stable neutron star. Of course, enough electrons and protons
must remain in the neutron star so that Pauli's exclusion principle prevents neutron
beta decay

\begin{center}
$n \rightarrow p + e^{-}  + \overline \nu_{e}$
\end{center}

Where $\overline \nu_{e}$ is the electron antineutrino (Weinberg 1972a). This requirement sets a lower limit $\sim 0.2 \msol$ on the mass of a stable
neutron star.\\
If,  however, after the end of the thermonuclear evolution, the mass of the core
of a star is greater than the Chandrasekhar and Oppenheimer-Volkoff limit, the star
may eject enough matter so that the mass of the core drops below the Chandrasekhar
and Oppenheimer-Volkoff limit as a result of which it may settle as a stable white
dwarf or a stable neutron star. If not, the core will gravitationally collapse and
end up as a black hole.

As is well known, the event horizon of a black hole of mass $M$ is a spherical surface located
at a distance $r = r_{g} = 2GM/c^{2}$ from the centre, where $G$ is Newton's
gravitational constant and $c$ the speed of light in vacuum; $r_{g}$ is called
gravitational radius or Schwarzschild radius. An external observer cannot observe
anything that is happening inside the event horizon, nothing, not even light or any
other electromagnetic signal can escape outside the event horizon from inside. However,
anything that enters the event horizon from outside is swallowed by the black hole;
it can never escape outside the event horizon again. 

Attempts have been made, using the general theory of relativity (GTR), to understand
what happens inside a black hole. In so doing, various simplifying assumptions have
been made. In the simplest treatment (Oppenheimer and Snyder 1939; Weinberg 1972b) a
black hole is considered to be a ball of dust with negligible pressure, uniform
density $\rho = \rho(t)$, and at rest at $t=0$. These assumptions lead to the unique
solution of the Einstein field equations, and in the comoving co-ordinates the metric
inside the black hole is given by

\begin{eqnarray}
ds^2 = dt^2 -R^2(t)\bbs \frac{dr^2}{1-k\,r^2} + r^2 d\theta^2 + r^2\sin^2\theta\,d\phi^2\ebs
\end{eqnarray}
in units in which speed of light in vacuum, c=1, and where $k$ is a constant. The requirement 
of energy conservation implies that $\rho(t)R^3(t)$ remains constant. On normalizing the 
radial co-ordinate $r$ so that

\begin{eqnarray}
R(0) = 1
\end{eqnarray}
one gets

\begin{eqnarray}
\rho(t) = \rho(0)R^{-3}(t)
\end{eqnarray}
The fluid is assumed to be at rest at $t=0$, so

\begin{eqnarray}
\dot{R}(0) = 0
\end{eqnarray}
Consequently, the field equations give

\begin{eqnarray}
 k = \frac{8\pi\,G}{3} \rho(0)
\end{eqnarray}

Finally, the solution of the field equations is given by the parametric
equations of a cycloid :

\begin{eqnarray}
\nonumber
t = \bb \frac{\psi + \sin\,\psi}{2\sqrt{k}} \eb \\
R = \frac{1}{2} \bb 1 + \cos\,\psi \eb
\end{eqnarray}

From equation $(6)$ it is obvious that when $\psi = \pi $. i.e., when

\begin{eqnarray}
t = t_{s} = \frac{\pi}{2\sqrt{k}} = \frac{\pi}{2} \bb \frac{3}{8\pi\,G \rho(0)} \eb^{1/2}
\end{eqnarray}
a space-time singularity occurs; the scale factor $R(t)$ vanishes. In other words,
a black hole of uniform density having the initial values $\rho(0)$, and zero
pressure collapses from rest to a point in $3$ - subspace, i.e., to a $3$ - subspace
of infinite curvature and zero proper volume, in a finite time $t_{s}$; the collapsed
state being a state of infinite proper energy density. The same result is obtained
in the Newtonian collapse of a ball of dust under the same set of assumptions
(Narlikar 1978).

Although the black hole collapses completely to a point at a finite co-ordinate time
$t=t_{s}$, any electromagnetic signal coming to an observer on the earth from the surface
 of the collapsing star before it crosses its event horizon will be delayed by its
gravitational field, so an observer on the earth will not see the star suddenly vanish.
Actually, the collapse to the  Schwarzschild radius $r_{g}$ appears to an outside
observer to take an infinite time, and the collapse to $R=0$ is not at all observable
from outside the event horizon.

The internal dynamics of a non-idealized, real black hole is very complex. Even in
the case of a spherically symmetric collapsing black hole with non-zero pressure
the details of the interior dynamics are not well understood, though major advances
in the understanding of the interior dynamics are now being made by means of
numerical computations and analytic analyses. But in these computations and analyses
no new features have emerged beyond those that occur in the simple uniform-density,
free-fall collapse considered above (Misner,Thorne, and Wheeler 1973). However,
using topological methods, Penrose (1965,1969), Hawking (1996a, 1966b, 1967a, 1967b),
Hawking and Penrose (1970), and Geroch (1966, 1967, 1968) have proved a number of
singularity theorems purporting that if an object contracts to dimensions smaller
than $r_{g}$, and if other reasonable conditions - namely, validity of the GTR,
positivity of energy, ubiquity of matter and causality - are satisfied, its collapse
to a singularity is inevitable.
\section{A critique of the singularity theorems}

As mentioned above, the singularity theorems are based, inter alia, on the assumption that the
GTR is universally valid. But the question is : Has the validity of the GTR been established
experimentally in the case of strong fields ? Actually, the GTR has been experimentally verified
only in the limiting case of week fields, it has not been experimentally validated in the case
of strong fields. Moreover, it has been demonstrated that when curvatures exceed the critical
value $C_{g} = 1/L_{g}^4$, where $L_{g} =  \bb \hbar\,G/c^{3} \eb^{1/2} = 1.6 \times 10^{-33}
\cm $ corresponding to the critical density $\rho_{g} = 5 \times 10^{93} \gcmcui$, the GTR is
no longer valid; quantum effects must enter the picture (Zeldovich and Novikov 1971). Therefore,
 it is clear that the GTR breaks down before a gravitationally collapsing object collapses to
 a singularity. Consequently, the conclusion based on the GTR that in comoving co-ordinates any
  gravitationally collapsing object in general, and a black hole in particular, collapses to a
  point in 3-space need not be held sacrosanct, as a matter of fact it may not be correct at
  all.

Furthermore, while arriving at the singularity theorems attention has mostly been focused on the
 space-time geometry and geometrodynamics; matter has been tacitly treated as a classical
 entity. However, as will be shown later, this is not justified; quantum mechanical behavior of
 matter at high energies and high densities must be taken into account. Even if we regard
 matter as a classical entity of a sort, it can be easily seen that the collapse of a black
 hole to a space-time singularity is inhibited by Pauli's exclusion principle. As mentioned
 earlier, a collapsing black hole consists, almost entirely, of neutrons apart from traces of
 protons and electrons; and neutrons  as well as protons and electrons are fermions; they obey
 Pauli's exclusion principle. If a black hole collapses to a point in 3-space, all the neutrons
 in the black hole would be squeezed into just two quantum states available at that point, one
 for spin up and the other for spin down neutron. This would violate Pauli's exclusion
 principle, according to which not more than one fermion of a given species can occupy any
 quantum state. So would be the case with the protons and the electrons in the black hole.
 Consequently, a black hole cannot collapse to a space-time singularity in contravention to
 Pauli's exclusion principle.

Besides, another valid question is : What happens to a black hole after $t > t_{s}$, i.e., after it
has collapsed to a point in 3-space to a state of infinite proper energy density, if at all such
 a collapse occurs? Will it remain frozen forever at that point? If yes, then uncertainties in
 the position co-ordinates of each of the particles - namely, neutrons, protons, and electrons -
 comprising the black hole would be zero. Consequently, according to Heisenberg's uncertainty
 principle, uncertainties in the momentum co-ordinates of each of the particles would be infinite.
  However, it is physically inconceivable how particles of infinite momentum and energy would
  remain frozen forever at a point. From this consideration also collapse of a black hole to a
  singularity appears to be quite unlikely.

Earlier, it was suggested by the author that the very strong 'hard-core' repulsive interaction
between nucleons, which has a range $l_{c} \sim 0.4 \times 10^{-13} \cm $, might set a limit on
the gravitational collapse of a black hole and avert its collapse to a singularity (Thakur 1983).
 The existence of this hard-core interaction was pointed out by Jastro (1951) after the analysis
  of the data from high energy nucleon-nucleon scattering experiments. It has been shown that 
  this very strong short
  range repulsive interaction arises due to the exchange of isoscalar vector mesons $\omega$ and
   $\phi$ between two nucleons ( Scotti and Wong 1965). Phenomenologically, that part of the
   nucleon-nucleon potential which corresponds to the repulsive hard core interaction may be
   taken as

\begin{eqnarray}
V_{c}(r) = \infty~~~~~~~~~for~~~  r < l_{c}
\end{eqnarray}
where $r$ is the distance between the two interacting nucleons. Taking
this into account, the author concluded that no spherical object of
mass M could collapse to a sphere of radius smaller than $R_{min} =
1.68 \times 10^{-6} M^{1/3} \cm$, or of the density greater than
$\rho_{max} = 5.0 \times 10^{16} \gcmcui$. It was also pointed out
that an object of mass smaller than $M_{c} \sim 1.21 \times 10^{33}
\gm$ could not cross the event horizon and become a black hole; the
only course left to an object of mass smaller than $M_{c}$ was to
reach equilibrium as either a white dwarf or a neutron star. However,
one may not regard these conclusions as reliable because they are based
on the hard core repulsive interaction (8) between nucleons which has
been arrived at phenomenologically by high energy nuclear physicists
while accounting for the high energy nucleon-nucleon scattering data;
but it must be noted that, as mentioned above, the existence of the hard
core interaction has been demonstrated theoretically also by Scotti
and Wong in 1965. Moreover, it is interesting to note that the upper
limit $M_{c} \sim 1.21 \times 10^{33} \g = 0.69 \msol$ on the masses
of objects that cannot gravitationally collapse to form black holes is
of the same order of magnitude as the Chandrasekhar and the
Oppenheimer- Volkoff limits.

Even if we disregard the role of the hard core, short range repulsive
interaction in arresting the collapse of a black hole to a space-time
singularity in comoving co-ordinates, it must be noted that unlike
leptons which appear to be point-like particles - the experimental
upper bound on their radii being $10^{-16} \cm$ (Barber \etal 1979)
-nucleons have finite dimensions. It has been experimentally
demonstrated that the radius $r_0$ of the proton is about $10^{-13}
\cm$(Hofstadter \& McAllister 1955). Therefore, it is natural to assume 
that the radius $r_0$ of the neutron is also about $10^{-13} \cm$.  This 
means the minimum volume $v_{min}$ occupied by a neutron is
$\frac{4\pi}{3}{r_0}^3$. Ignoring the ``mass defect'' arising from
the release of energy during the gravitational contraction (before
crossing the event horizon), the number of neutrons $N$ in a
collapsing black hole of mass $M$ is, obviously, $\frac{M}{m_{n}}$
where $m_{n}$ is the mass of the neutron. Assuming that neutrons are
impregnable particles, the minimum volume that the black hole can
occupy is $V_{min} = Nv_{min} = v_{min} \frac{M}{m_{n}} $, for neutrons cannot
be more closely packed than this in a black hole. However,
$V_{min} = \frac{4\pi R_{min}^3}{3}$ where $R_{min}$ is the radius of
the minimum volume to which the black hole can
collapse. Consequently, $R_{min} = r_{0} {\bb\frac{M}{m_{n}}\eb}^{1/3}$. On
substituting $10^{-13} \cm$ for $r_{0}$ and $1.67 \times 10^{-24} \g$ for
$m_n$ one finds that $R_{min} = 8.40 \times 10^{-6} M^{1/3}$.
This means a collapsing black hole cannot collapse to a
density greater than $\rho_{max} = \frac{M}{V_{min}} =
\frac{Nm_{n}}{4/3 \pi r_{0}^{3} N} = 3.99 \times 10^{14}
\gcmcui$. The critical mass $M_{c}$ of the object for which the
gravitational radius $R_{g} = R_{min}$ is obtained from the equation
\begin{eqnarray}
\frac{2GM_{c}}{c^{2}} = r_{0} \bb \frac{M_{c}}{m_{n}} \eb^{1/3}
\end{eqnarray}
This gives
\begin{eqnarray}
M_{c} = 1.35 \times 10^{34} \g = 8.68 \msol
\end{eqnarray}
Obviously, for $M > M_{c}, R_{g} > R_{min} $, and for $M < M_{c},
R_{g} < R_{min} $.

Consequently, objects of mass $M < M_{c}$ cannot cross the event
 horizon and become a black hole whereas those of mass $M > M_{c}$
 can. Objects of mass $M < M_{c}$ will, depending on their mass, reach
 equilibrium as either white dwarfs or neutron stars. Of course,
 these conclusions are based on the assumption that neutrons are
 impregnable particles and have radius $r_{0} = 10^{-13} cm $
 each. Also implicit is the assumption that neutrons are {\it
 fundamental } particles; they are not composite particles made up of
 other smaller constituents. But this assumption is not correct;
 neutrons as well as protons and other hadrons are {\it not
 fundamental} particles; they are made up of smaller constituents
 called {\it quarks } as will be explained in section 4. In section 5
 it will be shown how, at ultrahigh energy and ultrahigh density,
 the entire matter in a collapsing black hole is eventually converted
 into quark-gluon plasma permeated by leptons.
\section{Gravitationally collapsing black hole as a particle accelerator}
We consider a gravitationally collapsing black hole. On neglecting mutual interactions the
energy E of any one of the particles comprising the black hole is given by $E^2 = p^2 + m^2 >
p^2$, in units in which the speed of light in vacuum $c= 1$, where $p$ is the magnitude of the
3-momentum of the particle and $m$ its rest mass. But $p = \frac{h}{\lambda}$, where
$\lambda$ is the de Broglie wavelength of the particle and $h$ Planck's constant of action.
Since all lengths in the collapsing black hole scale down in proportion to the scale factor
$R(t)$ in equation $(1)$, it is obvious that $\lambda \propto R(t) $. Therefore it follows
that $p \propto R^{-1}(t)$, and hence $p = a\,R^{-1}(t)$, where {\it a} is the constant of
proportionality. From this it follows that $E > a/R$. Consequently, $E$ as well as $p$
 increases continually as R decreases. It is also obvious that $E$ and $p$, the magnitude of the
 3-momentum, $\rightarrow \infty$ as $R \rightarrow 0$. Thus, in effect, we have an
 ultra-high energy particle accelerator, {\it so far inconceivable in any terrestrial
 laboratory}, in the form of a collapsing black hole, which can, in the absence of any physical
 process inhibiting the collapse, accelerate particles to an arbitrarily high energy and momentum
 without any limit.

What has been concluded above can also be demonstrated alternatively, without
resorting to GTR, as follows. As an object collapses under its selfgravitation, the
interparticle distance $s$ between any pair of particles in the object decreases. Obviously,
the de Broglie's wavelength $\lambda$ of any particle in the object is less than or equal to $s$,
a simple consequence of Heisenberg's uncertainty principle. Therefore, $s \geq h/p $,
where $h$ is Planck's constant and $p$ the magnitude of 3-momentum of the particle.
Consequently, $p \geq h/s $ and hence $E \geq h/s $. Since during the collapse of the
 object $s$ decreases, the energy $E$ as well as the momentum $p$ of each of the particles in
  the object increases. Moreover, from $E \geq h/s $ and $p \geq h/s $ it follows
   that $E$ and $p \rightarrow \infty$ as $ s \rightarrow 0$. Thus, any gravitationally
   collapsing object in general, and a black hole in particular, acts as an ultrahigh energy
   particle accelerator.

It is also obvious that $\rho$, the density of matter in the black hole, increases as it
collapses. In fact, $\rho \propto R^{-3} $, and hence $\rho \rightarrow \infty$ as 
$R \rightarrow 0$.
\section{Quarks: The building blocks of matter}
In order to understand eventually what happens to matter in a collapsing black hole one has to
take into account the microscopic behavior of matter at high energies and high densities;
one has to consider the role played by the electromagnetic, weak, and strong interactions -
apart from the gravitational interaction - between the particles comprising the matter. For a
brief account of this the reader is referred to Thakur(1995), for greater detail to Huang(1992),
 or at a more elementary level to Hughes(1991).

As has been mentioned in Section 2, unlike leptons, hadrons are not point-like particles, but
are of finite size; they have structures which have been revealed in experiments that probe
hadronic structures by means of electromagnetic and weak interactions. The discovery of a very
large number of {\it apparently elementary (fundamental)} hadrons led to the search for a
pattern amongst them with a view to understanding their nature. This resulted in attempts to
group together hadrons having the same baryon number, spin, and parity but different strangeness
 $S$ ( or equivalently hypercharge $Y = B + S$, where $B$ is the baryon number) into I-spin
 (isospin) multiplets. In a plot of $Y$ against $I_{3}$ (z- component of isospin I), members
 of I-spin  multiplets are represented by points. The existence of several such hadron
 (baryon and meson) multiplets is a manifestation of underlying internal symmetries.

In 1961 Gell-Mann, and independently Ne\'emann, pointed out that each of these multiplets can be
looked upon as the realization of an irreducible representation of an internal symmetry group
$SU(3)$ ( Gell-Mann and Ne\'emann 1964). This fact together with the fact that hadrons have finite
size and inner structure led Gell-Mann, and independently Zweig, in 1964 to hypothesize that
hadrons {\it are not elementary particles}, rather they are composed of more elementary
constituents called {\it quarks ($q$)} by Gell-Mann (Zweig called them {\it aces}).
Baryons are composed of three quarks ($q\,q\,q$) and antibaryons of three antiquarks
($\overline q\,\overline q\, \overline q$) while mesons are composed of a quark and an
antiquark each. In the beginning, to account for the multiplets of baryons and mesons, quarks of
only three flavours, namely, u(up), d (down), and s(strange) were postulated, and they together
formed the basic triplet  $\left( \begin{array}{c}u\\d\\  s \end{array} \right)$ of the internal
 symmetry group $SU(3)$. All these three quarks u, d, and s have spin 1/2 and baryon number 1/3.
  The u quark has charge $2/3\,e$  whereas the d and s quarks have charge $-1/3\,e$ where $e$
   is the charge of the proton. The strangeness quantum number of the u and d quarks is zero
   whereas that of the s quark is -1. The antiquarks ($\overline u\,,\overline d\,,\overline s$)
    have charges $-2/3\,e, 1/3\,e,1/3\,e$ and strangeness quantum numbers 0, 0, 1 respectively.
    They all have spin 1/2 and baryon number -1/3. Both u and d quarks have the same mass,
    namely, one third that of the nucleon, i.e., $\simeq 310 MeV/c^2 $ whereas the mass of the $s$
    quark is $\simeq 500 MeV/c^2$. The  proton is composed of two up and one down quarks
    (p: uud) and the neutron of one up and two down quarks (n: udd).

Motivated by certain theoretical considerations Glashow, Iliopoulos and Maiani (1970) proposed
that, in addition to $u$, $d$, $s$ quarks, there should be another quark flavour which they
named {\it charm} $(c)$. Gaillard and Lee (1974) estimated its mass to be $\simeq 1.5 GeV/c^{2}$.
In 1974 two teams, one led by S.C.C. Ting at SLAC (Aubert \etal 1974) and another led by B.
Richter at Brookhaven (Augustin \etal 1974) independently discovered the $J/\Psi$, a particle
remarkable in that its mass ($3.1 GeV/c^{2}$) is more than three times that of the proton.
Since then, four more particles of the same family, namely, $\psi (3684)$, $\psi (3950)$, $\psi
(4150)$, $\psi (4400)$ have been found. It is now established that these particles are bound
states of {\it charmonium} ($\overline c c$), $J/\psi$ being the ground state. On adopting
non-relativistic independent quark model with a linear potential between $c$ and $\overline c$,
 and taking the mass of $c$ to be approximately half the mass of $J/\psi$, \ie $1.5 GeV/c^{2}$,
  one can account for the  $J/\psi$ family of particles. The $c$ has spin $1/2$, charge $2/3$
  $e$, baryon number $1/3$, strangeness $-1$, and a new quantum number charm $(c)$ equal to 1.
  The $u$, $d$, $s$ quarks have $c=0$. It may be pointed out here that charmed mesons and
   baryons, \ie the bound states like ($c\overline d$), and ($cdu$) have also been found. Thus
   the existence of the $c$ quark has been established experimentally beyond any shade of
   doubt.

The discovery of the $c$ quark stimulated the search for more new quarks. An additional
motivation for such a search was provided by the fact that there are three generations of
lepton {\it weak} {\it doublets}: ${\nu_{e}\choose e}$, ${\nu_{\mu}\choose \mu},$ and
${\nu_{\tau}\choose \tau}$ where $\nu_{e}$, $\nu_{\mu},$ and $\nu_{\tau}$ are electron ($e$),
muon ($\mu$), and tau lepton ($\tau$) neutrinos respectively. Hence, by analogy, one expects
that there should be three generations of quark {\it weak} {\it doublets} also: ${u \choose d}$,
  ${c \choose s}$, and  ${? \choose ?}$. It may be mentioned here that weak interaction does
  not distinguish between the upper and the lower members of each of these doublets. In
  analogy with the isopin $1/2$ of the {\it strong doublet}  ${p \choose n}$, the {\it weak
  doublets} are regarded as possessing {\it weak isopin} $I_{W} = 1/2$, the third component
  $(I_{W})_{3}$ of this {\it weak isopin} being + 1/2 for the upper components of these
  doublets and - 1/2 for the lower components. These statements apply to the left-handed
   quarks and leptons, \ie those with negative helicity (\ie with the spin antiparallel to
   the momentum) only. The right-handed leptons and quarks, \ie those with positive
   helicity (\ie with the spin parallel to the momentum), are {\it weak singlets} having
   {\it weak isopin} zero.

The discovery, at Fermi Laboratory, of a new family of vector mesons, the upsilon family,
starting at a mass of $9.4 GeV/c^{2}$ gave an evidence for a new quark flavour called
{\it bottom} or {\it beauty} $(b)$ (Herb \etal 1997; Innes \etal 1977). These vector
mesons are in fact, bound states of bottomonium $(\overline b b)$. These states have since
been studied in detail at the Cornell electron accelerator in an electron-positron storage
ring of energy ideally matched to this mass range. Four such states with masses $9.46,
10.02, 10.35,$ and $10.58$ $GeV/c^{2}$ have been found, the state with mass $9.46 GeV/c^{2}$
 being the ground state (Andrews \etal 1980). This implies that the mass of the $b$
 quark is $\simeq 4.73 GeV/c^{2}$. The $b$ quark has spin $1/2$ and charge $-1/3\ e$.
 Furthermore, the $b$ flavoured mesons have been found with exactly the expected properties
  (Beherend \etal 1983).

After the discovery of the $b$ quark, the confidence in the existence
of the sixth quark flavour called {\it top} or {\it truth} $(t)$
increased and it became almost certain that, like leptons, the quarks
also occur in three generations of weak isopin doublets, namely, ${u
\choose d}$, ${c \choose s}$, and ${t\choose b}$. In view of this,
intensive search was made for the $t$ quark. But the discovery of the
$t$ quark eluded for eighteen years.  However, eventually in 1995, two
groups, the CDF (Collider Detector at Fermi lab) Collaboration (Abe
\etal 1995) and the $D\phi$ Collaboration (Abachi \etal 1995)
succeeded in detecting {\it toponium} $\overline t t$ in very high
energy $\overline p p$ collisions at Fermi Laboratory's $1.8
TeV$Tevetron collider. The {\it toponium} $\overline t t$ is the bound
state of $t$ and $\overline t$. The mass of $t$ has been estimated to
be $176.0\pm2.0 GeV/c^{2}$, and thus it is the most massive elementary
particle known so far. The $t$ quark has spin $1/2$ and charge $2/3\
e$.

Moreover, in order to account for the apparent breaking of the
spin-statistics theorem in certain members of the
$J^{p}=\frac{3^{+}}{2}$ decuplet (spin 3/2,parity even), \eg,
$\bigtriangleup^{++}$ $(uuu)$, and $\Omega^{-}$ $(sss)$, Greenberg
$(1964)$ postulated that quark of each flavour comes in three {\it
colours}, namely, {\it red}, {\it green}, and {\it blue}, and that real
particles are always {\it colour singlets}. This implies that real
particles must contain quarks of all the three colours or
colour-anticolour combinations such that they are overall {\it white} or
{\it colourless}. {\it White} or {\it colourless} means all the three
primary colours are equally mixed or there should be a combination of a
quark of a given colour and an antiquark of the corresponding
anticolour. This means each baryon contains quarks of all the three colours(but not 
necessarily of the same flavour) whereas a meson contains a quark of a given 
colour and an antiquark having the corresponding anticolour so that each combination 
is overall white. Leptons have no colour. Of course, in this context the word
`colour' has nothing to do with the actual visual colour, it is just a
quantum number specifying a new internal degree of freedom of a quark.

The concept of colour plays a fundamental role in accounting for the
interaction between quarks. The remarkable success of quantum
electrodynamics (QED) in explaining the interaction between electric
charges to an extremely high degree of precision motivated physicists
to explore a similar theory for strong interaction. The result is
quantum chromodynamics (QCD), a non-Abelian gauge theory (Yang-Mills
theory), which closely parallels QED. Drawing analogy from
electrodynamics, Nambu (1966) postulated that the three quark colours
are the charges (the Yang-Mills charges) responsible for the force
between quarks just as electric charges are responsible for the
electromagnetic force between charged particles. The analogue of the
rule that like charges repel and unlike charges attract each other is
the rule that like colours repel, and colour and anticolour attract each
other. Apart from this, there is another rule in QCD which states that
different colours attract if the quantum state is antisymmetric, and
repel if it is symmetric under exchange of quarks. An important
consequence of this is that if we take three possible pairs,
red-green. green-blue, and blue-red, then a third quark is attracted
only if its colour is different and if the quantum state of the resulting combination 
is antisymmetric under the exchange of a pair of quarks thus resulting in
red-green-blue baryons. Another consequence of this rule is that a
fourth quark is repelled by one quark of the same colour and attracted by two of different colours
 in a baryon but only in antisymmetric combinations. This introduces a factor of
1/2 in the attractive component and as such the overall force is zero,
i.e., the fourth quark is neither attracted nor repelled by a
combination of red-green-blue quarks. In spite of the fact that
hadrons are overall colourless, they feel a residual strong force due
to their coloured constituents.

 It was soon realized that if the three colours are to serve as the
 Yang-Mills charges, each quark flavour must transform as a triplet of
 $SU_{c}(3)$ that causes transitions between quarks of the same flavour
 but of different colours ( the SU(3) mentioned earlier causes
 transitions between quarks of different flavours and hence may more
 appropriately be denoted by $SU_{f}(3)$). However, the $SU_{c}(3)$
 Yang-Mills theory requires the introduction of eight new spin 1 gauge
 bosons called {\it gluons}. Moreover, it is reasonable to stipulate
 that the gluons couple to {\it left-handed} and {\it right-handed}
 quarks in the same manner since the strong interactions do not
 violate the law of conservation of parity. Just as the force between
 electric charges arise due to the exchange of a photon, a massless
 vector (spin 1) boson, the force between coloured quarks arises due to
 the exchange of a gluon. Gluons are also massless vector (spin 1)
 bosons. A quark may change its colour by emitting a gluon. For
 example, a {\it red} quark $q_{R}$ may change to a blue quark $q_{B}$
 by emitting a gluon which may be thought to have taken away the {\it
 red (R) colour } from the quark and given it the {\it blue (B)} colour,
 or, equivalently, the gluon may be thought to have taken away the
 {\it red (R)} and the {\it antiblue ($\overline B$)} colours from the
 quark. Consequently, the {\it gluon $G_{RB}$} emitted in the process
 $q_{R} \rightarrow q_{B}$ may be regarded as the composite having the
 {\it colour R $\overline B$ } so that the emitted gluon $G_{RB} =
 q_{R}\overline q_{B}$. In general, when a quark $q_{i}$ of {\it colour
 i} changes to a quark $q_{j}$ of {\it colour j} by emitting a gluon $G_{ij}$,
 then $G_{ij}$ is the composite state of $q_{i}$ and $\overline
 q_{j}$, i.e., $G_{ij} = q_{i} \overline q_{j}$. Since there are three
 {\it colours} and three{\it anticolours}, there are $3 \times 3 = 9$
 possible combinations ({\it gluons})of the form $G_{ij} = q_{i}
 \overline q_{j}$. However, one of the nine combinations is a special
 combination corresponding to the {\it white colour}, namely, $G_{W} =
 q_{R} \overline q_{R} = q_{G} \overline q_{G} = q_{B} \overline
 q_{B}$. But there is no interaction between a {\it coloured} object
 and a {\it white (colourless)} object. Consequently, gluon $G_{W}$
 may be thought not to exist. This leads to the conclusion that only
 $9 - 1 = 8$ kinds of gluons exist. This is a heuristic explanation of
 the fact that $SU_{c}(3)$ Yang-Mills gauge theory requires the
 existence of eight gauge bosons, i.e., the gluons. Moreover, as the
 gluons themselves carry colour, gluons may also emit gluons. Another
 important consequence of gluons possessing colour is that several
 gluons may come together and form {\it gluonium} or {\it glue
 balls}. Glueballs have integral spin and no colour and as such they
 belong to the meson family.

    Though the actual existence of quarks has been indirectly
    confirmed by  experiments that probe hardronic structure by means
    of electromagnetic and weak interactions, and by the production
    of various quarkonia ($\overline q q $) in high energy collisions
    made possible by various particle accelerators, no {\it free}
    quark has been detected in experiments at these accelerators so
    far. This fact has been attributed to the {\it infrared slavery}
    of quarks, i.e., to the nature of the interaction between quarks
    responsible for their {\it confinement} inside hadrons. Perhaps
    enormous amount of energy , much more than what is available in
    the existing terrestrial accelerators, is required to liberate the
    quarks from confinement. This means the force of attraction
    between quarks increases with increase in their separation. This
    is reminiscent of the force between two bodies connected by
    an elastic string.

    On the contrary, the results of deep inelastic scattering
    experiments reveal an altogether different feature of the
    interaction between quarks. If one examines quarks at very short
    distances ($ < 10^{-13}$ cm ) by observing the scattering of a
    nonhadronic probe, e.g., an electron or a neutrino, one finds that
    quarks move almost freely inside baryons and mesons as though they
    are not bound at all. This phenomenon is called the {\it
    asymptotic freedom} of quarks. In fact Gross and Wilczek (1973
    a,b) and Politzer (1973) have shown that the running coupling
    constant of interaction between two quarks vanishes in the limit
    of infinite momentum (or equivalently in the limit of zero
    separation).
\section{Eventually what happens to matter in a collapsing black hole?}

    As mentioned in Section 3 the energy $E$ of the particles
    comprising the matter in a collapsing black hole continually
    increases and so does the density $\rho$ of the matter whereas the
    separation $s$ between any pair of particles decreases. During the
    continual collapse of the black hole a stage will be reached when
    $E$ and $\rho$ will be so large and $s$ so small that the quarks
    confined in the hadrons will be liberated from the {\it infrared
    slavery} and will enjoy {\it asymptotic freedom}, i.e., the quark
    {\it deconfinement} will occur. In fact, it has been shown that
    when the energy $E$ of the particle $\sim 10^{2}$ GeV ($s \sim
    10^{-16}$ cm) corresponding to a temperature $T \sim 10^{15} K$
    all interactions are of the Yang-Mills type with $SU_{c}(3) \times
    SU_{I_W}(2) \times U_{Y_W}(1)$ gauge symmetry, where c stands for
    colour, $I_{W}$ for weak isospin, and $Y_{W}$ for weak hypercharge,
    and at this stage quark deconfinement occurs as a result of which
    matter now consists of its fundamental constituents : spin 1/2
    leptons, namely, the electrons, the muons, the tau leptons, and their
    neutrinos, which interact only through the electroweak
    interaction(i.e., the unified electromagnetic and weak
    interactions); and the spin 1/2 quarks, u, d, s, c, b, t, which
    interact eletroweakly as well as through the colour force generated
    by gluons(Ramond, 1983). In other words, when $E \geq 10^{2}$ GeV
    ($s \leq 10^{-16}$ cm) corresponding to $T \geq 10^{15} K$, the
    entire matter in the collapsing black hole will be in the form of
    qurak-gluon plasma permeated by leptons as suggested by the author
    earlier (Thakur 1993).

    Incidentally, it may be mentioned that efforts are being made to
    create quark-gluon plasma in terrestrial laboratories. A report
    released by CERN, the European Organization for Nuclear Research,
    at Geneva, on February 10, 2000, said that by smashing together lead ions at CERN's
    accelerator at temperatures 100,000 times as hot as the Sun's
    centre, i.e., at $T \sim 1.5 \times 10^{12} K$, and  energy
    densities never before reached in laboratory experiments, a team
    of 350 scientists from institutes in 20 countries succeeded in
    isolating tiny components called quarks from more complex
    particles such as protons and neutrons. ``A series of experiments
    using CERN's lead beam have presented compelling evidence for the
    existence of a new state of matter 20 times denser than nuclear
    matter, in which quarks instead of being bound up into more complex
    particles such as protons and neutrons, are 
    liberated to roam freely '' the report said. However, the evidence
    of the creation of quark gluon plasma at CERN is indirect,
    involving detection of particles produced when the quark-gluon
    plasma changes back to hadrons. The production of these
    particles can be explained alternatively without having to have
    quark-gluon plasma. Therefore, Ulrich Heinz at CERN is of the
    opinion that the evidence of the creation of quark-gluon plasma at
    CERN is not enough and conclusive. In view of this, CERN will start a new experiment, ALICE, soon (around 2007-2008) at its Large Hadron Collider (LHC) in 
    order to definitively and conclusively creat QGP.

    In the meantime the focus of research on quark-gluon plasma has
    shifted to the Relativistic Heavy Ion Collider (RHIC), the worlds
    newest and largest particle accelerator for nuclear research, at
    Brookhaven National Laboratory in Upton, New York. RHIC's goal is
    to create and study quark-gluon plasma. RHIC's aim is to create
    quark-gluon plasma by head-on collisions of two beams of gold ions
    at energies 10 times those of CERN's programme, which ought to
    produce a quark-gluon plasma with higher temperature and longer
    lifetime thereby allowing much clearer and direct
    observation. RHIC's quark-gluon plasma is expected to be well
    above the transition temperature for transition between the
    ordinary hadronic matter phase and the quark-gluon plasma
    phase. This will enable scientists to perform numerous advanced
    experiments in order to study the properties of the plasma. The
    programme at RHIC began in the summer of 2000 and after two years
    Thomas Kirk, Brookhaven's Associate Laboratory Director for High
    Energy Nuclear Physics, remarked, ``It is too early to say that we
    have discovered the quark-gulon plasma, but not too early to mark
    the tantalizing hints of its existence.'' Other definitive
    evidence of quark-gluon plasma will come from experimental
    comparisons of the behavior in hot, dense nuclear matter with
    that in cold nuclear matter. In order to accomplish this, the next
    round of experimental measurements at RHIC will involve collisions
    between heavy ions and light ions, namely, between gold nuclei and
    deuterons.
 
Later, on June 18, 2003 a special scientific colloquium was held
 at Brcokhaven Natioal Laboratory (BNL) to discuss the latest findings at RHIC.
 At the colloquium, it was announced that in the detector system known as STAR ( Solenoidal Tracker AT RHIC ) head-on collision between two beams of gold nuclei of energies of 130 GeV per nuclei resulted in the phenomenon called ``jet quenching``. STAR as well as three other experiments at RHIC viz., PHENIX, BRAHMS, and
 PHOBOS, detected suppression of ``leading particles``, highly 
 energetic individual particles that emerge from nuclear fireballs, in gold-gold collisions. Jet quenching and leading particle suppression are signs
 of QGP formation. The findings of the STAR experiment were presented at the 
 BNL colloquium by Berkeley Laboratory's NSD ( Nuclear Science Division ) physicist Peter Jacobs. 
\section{Collapse of a black hole to a space-time singularity is inhibited by Pauli's exclusion principle}

As quarks and leptons in the quark-gluon plasma permeated by leptons
into which the entire matter in a collapsing black hole is eventually
converted are fermions, the collapse of a black hole to a space-time
singularity in a finite time in a comoving co-ordinate system, as
stipulated by the singularity theorems of Penrose, Hawking and Geroch,
is inhibited by Pauli's exclusion principle. For, if a black hole
collapses to a point in 3-space, all the quarks of a given flavour and
colour would be squeezed into just two quantum states available at that
point, one for spin up and the other for spin down quark of that
flavour and colour. This would violate Pauli's exclusion principle
according to which not more than one fermion of a given species can
occupy any quantum state. So would be the case with quarks of each
distinct combination of colour and flavour as well as with leptons of
each species, namely, $e, \mu, \tau, \nu_{e}, \nu_{\mu}$  and
$\nu_{\tau}$. Consequently, a black hole cannot collapse to a space-time
singularity in contravention to Pauli's exclusion principle. Then the
question arises : If a black hole does not collapse to a space-time
singularity, what is its ultimate fate? In section 7 three
possibilities have been suggested.
\section{Ultimately how does a black hole end up?}

The pressure $P$ inside a black hole is given by
\begin{eqnarray}
P = P_{r} + \sum_{i,j}P_{ij} + \sum_{k}P_{k} + \sum_{i,j} \overline P_{ij} + \sum_{k} \overline P_{k} 
\end{eqnarray}
where $P_{r}$ is the radiation pressure, $P_{ij}$ the pressure of the
relativistically degenerate quarks of the $i^{th}$ flavour and
$j^{th}$ colour, $P_{k}$ the pressure of the relativistically
degenerate leptons of the $k^{th}$ species, $\overline P_{ij}$ the
pressure of relativistically degenerate antiquarks of the $i^{th}$
flavour and $j^{th}$ colour, $\overline P_{k}$ that of the
relativistically degenerate antileptons of the $k^{th}$ species. In
equation (11) the summations over $i$ and $j$ extend over all the six
flavours and the three colours of quarks, and that over $k$ extend
over all the six species of leptons. However, calculation of these
pressures are prohibitively difficult for several reasons. For
example, the standard methods of statistical mechanics for
calculation of pressure and equation of state are applicable when the
system is in thermodynamics equilibrium and when its volume is very
large, so large that for practical purpose we may treat it as
infinite. Obviously, in a gravitationally collapsing black hole, the
photon, quark and lepton gases cannot be in thermodynamic equilibrium
nor can their volume be treated as infinite. Moreover, at ultrahigh
energies and densities, because of the $SU_{I_W}$(2) gauge symmetry, 
transitions between the upper and lower components of quark and lepton 
doublets occur very frequently. In addition to this, because of the 
$SU_{f}$(3)  and $SU_{c}$(3)
gauge symmetries transitions between quarks of different flavours and
colours also occur. Furthermore, pair production and pair annihilation
of quarks and leptons create additional complications. Apart from
these, various other nuclear reactions may as well
occur. Consequently, it is practically impossible to determine the
number density and hence the contribution to the overall pressure $P$ inside 
the black hole by any species of elementary particle in a collapsing
black hole when $E \geq 10^2$ Gev ($s \leq 10^{-16}$ cm), or
equivalently, $T \geq 10^{15} K$. However, it may not be unreasonable
to assume that, during the gravitational collapse, the pressure $P$
inside a black hole increases monotonically with the increase in the
density of matter $\rho$. Actually, it might be given by the polytrope,
 $P = k\rho^{\frac{(n+1)}{n}}$, where $K$ is a constant and $n$ is polytropic index. Consequently, $P \rightarrow \infty$ as $ \rho
\rightarrow \infty$, i.e., $P \rightarrow \infty$ as the scale factor
$R(t) \rightarrow 0$ (or equivalently $s \rightarrow 0$). In view of
this, there are three possible ways in which a black hole may end up.

\noindent 1. During the gravitational collapse of a black hole, at a
certain stage, the pressure $P$ may be enough to withstand the gravitational force
and the object may become
gravitationally stable. Since at this stage the object consists
entirely of quark-gluon plasma permeated by leptons, it means it
would end up as a stable quark star. Indeed, such a possibility seems
to exist. Recently, two teams - one led by David Helfand of Columbia
University, NewYork (Slane, Helfand, and Murray 2002) and another led by
Jeremy Drake of Harvard-Smithsonian Centre for Astrophysics,
Cambridge, Mass. USA (Drake \etal 2002) studied independently two
objects, 3C58 in Cassiopeia, and RXJ1856.5-3754 in Corona Australis
respectively by combining data from the NASA's Chandra X-ray
Observatory and the Hubble Space Telescope, that seemed, at first, to
be neutron stars, but, on closer look, each of these objects showed
evidence of being an even smaller and denser object, possibly a quark
star.

\noindent 2. Since the collapse of a black hole is inhibited by Pauli's
exclusion principle, it can  collapse only upto a certain minimum
radius, say, $r_{min}$. After this, because of the tremendous amount
of kinetic energy, it would bounce back and expand, but only upto the
event horizon, i.e., upto the gravitational (Schwarzschild ) radius
$r_g$ since, according to the GTR, it cannot cross the event
horizon. Thereafter it would  collapse again upto the radius
$r_{min}$ and then bounce back upto the radius $r_g$. This process of
collapse upto the radius $r_{min}$ and bounce upto the radius $r_g$
would occur repeatedly. In other words, the black hole would
continually pulsate radially between the radii $r_{min}$ and $r_g$ and
thus become a pulsating quark star. However, this pulsation would
cause periodic variations in the gravitational field outside the event
horizon and thus produce gravitational waves which would propagate
radially outwards in all directions from just outside the event
horizon. In this way the pulsating quark star would act as a source of
gravitational waves. The pulsation may take a very long time to damp
out since the energy of the quark star (black hole) cannot escape
outside the event horizon except via the gravitational radiation
produced outside the event horizon. However, gluons in the quark-gluon
plasma may also act as a damping agent. In the absence of damping,
which is quite unlikely, the black hole would end up as a perpetually
pulsating quark star.

\noindent 3. The third possibility is that eventually a black hole may
explode; a {\it mini bang} of a sort may occur, and it may, after the
explosion, expand beyond the event horizon though it has been
emphasized by Zeldovich and Novikov (1971) that after a collapsing
sphere's radius decreases to $r < r_g$ in a finite proper time, its
expansion into the external space from which the contraction
originated is impossible, even if the passage of matter through
infinite density is assumed.

Notwithstanding Zeldovich and Novikov's contention based on the very
concept of event horizon, a gravitationally collapsing black hole may
also explode by the very same mechanism by which the big bang
occurred, if indeed it did occur. This can be seen as follows. At the
present epoch the volume of the universe is $\sim 1.5 \times 10^{85}
\cm^3$ and the density of the galactic material throughout the universe
is $\sim 2 \times 10^{-31} \gcmcui$ (Allen 1973). Hence, a
conservative estimate of the mass of the universe is $\sim 1.5 \times
10^{85} \times 2 \times 10^{-31} \g = 3 \times 10^{54} \g $. However,
according to the big bang model, before the big bang, the entire matter
in the universe was contained in an {\it ylem} which occupied very
very small volume. The gravitational radius of the ylem of mass $3
\times 10^{54} g $ was $ 4.45 \times 10^{21} \km$ (it must have been
larger if the actual mass of the universe were taken into account
which is greater than $3 \times 10^{54} \g $). Obviously, the radius of
the ylem was many orders of magnitude smaller than its gravitational
radius, and yet the ylem exploded with a big bang, and in due course of
time crossed the event horizon and expanded beyond it upto the present
Hubble distance $c/H_0 \sim 1.5 \times 10^{23} \km$ where $c$ is the
speed of light in vacuum and $H_0$ the Hubble constant at the present
epoch. Consequently, if the ylem could explode in spite of Zeldovich
and Novikov's contention, a gravitationally collapsing black hole can
also explode, and in due course of time expand beyond the event
horizon. The origin of the big bang, i.e., the mechanism by which the
ylem exploded, is not definitively known. However, the author has,
earlier proposed a viable mechanism (Thakur 1992) based on
supersymmetry/supergravity. But supersymmetry/supergravity have not
yet been validated experimentally.
\section{Conclusion}
From the foregoing three inferences may be drawn. One, eventually the entire matter in a collapsing black hole is 
converted into quark-gluon plasma permeated by leptons. Two,
the collapse of a black hole to a space - time singularity
is inhibited by Pauli's exclusion principle. Three, ultimately a black hole may end up in one of the three possible ways suggested in section 7. 
\begin{acknowledgements}
The author thanks Professor S. K. Pandey, Co-ordinator, IUCAA
Reference Centre, School of Studies in Physics, Pt. Ravishankar Shukla
University, Raipur, for making available the facilities of the
Centre. He also thanks Sudhanshu Barway, Mousumi Das for typing the
manuscript.
\end{acknowledgements}


\begin{thebibliography}{}
\bibitem{}Abachi S., \etal, 1995, PRL, 74, 2632
\bibitem{}Abe F., \etal, 1995, PRL, 74, 2626
\bibitem{}Allen C. W., 1993, {\it \underline{Astrophysical Quantities}}, The Athlone Press, University of London, 293
\bibitem{}Andrew D., \etal, 1980, PRL, 44, 1108
\bibitem{}Aubert J. J., \etal, 1974, PRL, 33, 1404
\bibitem{}Augustin J. E., \etal, 1974, PRL, 33, 1406
\bibitem{}Barber D.P.,\etal,1979, PRL, 43, 1915
\bibitem{}Beherend S., \etal, 1983, PRL, 50, 881
\bibitem{}Drake J. \etal, 2002, ApJ, 572, 996
\bibitem{}Gaillard M. K., Lee B. W., 1974, PRD, 10, 897
\bibitem{}Gell-Mann M., N\'eeman Y., 1964, {\it The Eightfold Way},
W. A. Benjamin, NewYork
\bibitem{}Geroch R. P., 1966, PRL, 17, 445
\bibitem{}Geroch R. P., 1967, {\it Singularities in Spacetime of General Relativity : 
Their Defination, Existence and Local Characterization,}
Ph.D. Thesis, Princeton University
\bibitem{}Geroch, R. P., 1968, Ann. Phys., 48, 526
\bibitem{}Galshow S. L., Iliopoulos J., Maiani L., 1970, PRD,
2,1285
\bibitem{}Greenberg O. W., 1964, PRL, 13, 598
\bibitem{}Gross D. J., Wilczek F., 1973a, PRL, 30, 1343
\bibitem{}Gros, D. J., Wilczek F., 1973b, PRD, 8, 3633
\bibitem{}Hawking S. W., 1966a, Proc. Roy. Soc., 294A, 511
\bibitem{}Hawking S. W., 1966b, Proc. Roy. Soc., 295A, 490
\bibitem{}Hawking S. W., 1967a, Proc. Roy. Soc., 300A, 187
\bibitem{}Hawking S. W., 1967b, Proc. Roy. Soc., 308A, 433
\bibitem{}Hawking S. W., Penrose R., 1970, Proc. Roy. Soc., 314A,
529
\bibitem{}Herb S. W., \etal, 1977, PRL, 39, 252
\bibitem{}Hofstadter R., McAllister R. W., PR, 98, 217
\bibitem{}Huang K., 1982, {\it Quarks, Leptons and Gauge Fields},
World Scientific, Singapore
\bibitem{}Hughes I. S., 1991, {\it Elementry Particles}, Cambridge
Univ. Press, Cambridge
\bibitem{}Innes W. R., \etal, 1977, PRL, 39, 1240
\bibitem{}Jastrow R., 1951, PR, 81, 165
\bibitem{}Misner C. W., Thorne K. S., Wheeler J. A., 1973, Gravtitation, Freemon, NewYork, 857  
\bibitem{}Nambu Y., 1966, in A. de Shalit (Ed.), {\it Preludes in
Theoretical Physics}, North-Holland, Amsterdam
\bibitem{}Narlikar J. V., 1978,{\it Lectures on General Relativity
and Cosmology}, The MacMillan Company of India Limited, Bombay, 152
\bibitem{}Oppenheimer,J. R., Snyder H., 1939, PR 56, 455
\bibitem{}Penrose R., 1965, PRL, 14, 57
\bibitem{}Penrose R., 1969, {\it Riv. Nuoro Cimento}, 1, Numero
Speciale, 252 
\bibitem{}Politzer H. D., 1973, PRL, 30, 1346
\bibitem{}Ramond P., 1983, Ann. Rev. Nucl. Part. Sc., 33, 31
\bibitem{}Scotti A., Wong D. W., 1965, PR, 138B, 145
\bibitem{}Slane P. O., Helfand D. J., Murray, S. S., 2002, ApJL,
571, 45
\bibitem{}Thakur R. K., 1983, Ap\&SS, 91, 285
\bibitem{}Thakur R. K., 1992, Ap\&SS, 190, 281
\bibitem{}Thakur R. K., 1993, Ap\&SS, 199, 159
\bibitem{}Thakur R. K., 1995, {\it Space Science Reviews}, 73, 273
\bibitem{}Weinberg S., 1972a, {\it Gravitation and Cosmology}, John
Wiley \& Sons, New York, 318
\bibitem{}Weinberg S., 1972b, {\it Gravitation and Cosmology}, John
Wiley \& Sons, New York. 342-349
\bibitem{}Zeldovich Y. B., Novikov I. D., 1971, {\it Relativistic
Astrophysics}, Vol. I, University of Chicogo Press, Chicago,144-148
\bibitem{}Zweig G., 1964, Unpublished CERN Report

\end{thebibliography}


\end{document}